\documentclass[aps,prmaterials,reprint,superscriptaddress, longbibliography]{revtex4-1}
\usepackage{graphicx}
\usepackage{natbib}
\usepackage{siunitx}
\usepackage{textcomp}
\usepackage{nicefrac}

\DeclareSIUnit\inch{inch}
\DeclareSIUnit\oersted{Oe}
\DeclareSIUnit\unitformula{f.u.}

\begin{document}

\title{Molecular beam epitaxy of the half-Heusler antiferromagnet CuMnSb}

\author{L.~Scheffler}
\affiliation{Physikalisches Institut (EP3), Universit\"at W\"urzburg, 97074 W\"urzburg, Germany}
\affiliation{Institute for Topological Insulators,  Universit\"at W\"urzburg, 97074 W\"urzburg, Germany}
\author{K.~Gas}
\affiliation{Institute of Physics, Polish Academy of Sciences, Aleja Lotnikow 32/46, PL-02-668 Warszawa, Poland}
\author{S.~Banik}
\affiliation{Physikalisches Institut (EP3), Universit\"at W\"urzburg, 97074 W\"urzburg, Germany}
\affiliation{Institute for Topological Insulators,  Universit\"at W\"urzburg, 97074 W\"urzburg, Germany}
\author{M.~Kamp}
\affiliation{Physikalisches Institut and R\"ontgen Center for Complex Material Systems, Universit\"at W\"urzburg, 97074 W\"urzburg, Germany}
\author{J.~Knobel}
\affiliation{Physikalisches Institut (EP3), Universit\"at W\"urzburg, 97074 W\"urzburg, Germany}
\affiliation{Institute for Topological Insulators,  Universit\"at W\"urzburg, 97074 W\"urzburg, Germany}
\author{H.~Lin}
\affiliation{Physikalisches Institut (EP3), Universit\"at W\"urzburg, 97074 W\"urzburg, Germany}
\affiliation{Max-Planck-Institut f\"ur Chemische Physik fester Stoffe, 01187 Dresden, Germany}
\author{C.~Schumacher}
\affiliation{Physikalisches Institut (EP3), Universit\"at W\"urzburg, 97074 W\"urzburg, Germany}
\affiliation{Institute for Topological Insulators,  Universit\"at W\"urzburg, 97074 W\"urzburg, Germany}
\author{C.~Gould}
\affiliation{Physikalisches Institut (EP3), Universit\"at W\"urzburg, 97074 W\"urzburg, Germany}
\affiliation{Institute for Topological Insulators,  Universit\"at W\"urzburg, 97074 W\"urzburg, Germany}
\author{M.~Sawicki}
\affiliation{Institute of Physics, Polish Academy of Sciences, Aleja Lotnikow 32/46, PL-02-668 Warszawa, Poland}
\author{J.~Kleinlein}
\email{johannes.kleinlein@physik.uni-wuerzburg.de}
\affiliation{Physikalisches Institut (EP3), Universit\"at W\"urzburg, 97074 W\"urzburg, Germany}
\affiliation{Institute for Topological Insulators,  Universit\"at W\"urzburg, 97074 W\"urzburg, Germany}
\author{L.~W.~Molenkamp}
\affiliation{Physikalisches Institut (EP3), Universit\"at W\"urzburg, 97074 W\"urzburg, Germany}
\affiliation{Institute for Topological Insulators,  Universit\"at W\"urzburg, 97074 W\"urzburg, Germany}
\affiliation{Max-Planck-Institut f\"ur Chemische Physik fester Stoffe, 01187 Dresden, Germany}

\date{\today}

\begin{abstract}
We report growth of CuMnSb thin films by molecular beam epitaxy on InAs(001) substrates.
The CuMnSb layers are compressively strained (\SI{0.6}{\percent}) due to lattice mismatch.
The thin films have a $\omega$ full width half max of \SI{7.7}{\arcsecond} according to high resolution X-ray diffraction, and a root mean square roughness of \SI{0.14}{\nano\meter} as determined by atomic force microscopy.
Magnetic and electrical properties are found to be consistent with reported values from bulk samples.
We find a N\'{e}el temperature of \SI{62}{\kelvin}, a Curie-Weiss temperature of \SI{-65}{\kelvin} and an effective moment of \SI{5.9}{\mu_{B}/\unitformula{}}.
Transport measurements confirm the antiferromagetic transition and show a residual resistivity at \SI{4}{\kelvin} of \SI{35}{\micro\ohm\cdot\centi\meter}.

\end{abstract}
\maketitle

\section{Introduction}
Current trends in antiferromagnetic spintronics \cite{Jungwirth2018} are driving a demand for high quality materials in order to reliably study the phenomenology in this material class.
In particular, thin film specimens with a high degree of crystalline order and a low defect density are needed to access the fundamental mechanisms in transport experiments.
A promising antiferromagetic model system is the half-Heusler compound CuMnSb.
Although bulk samples of CuMnSb prepared by various melting techniques have been studied extensively over the past 50 years, the single-crystal thin films required for detailed transport experiments and device applications have yet to be realized.
The same is true for fabricating sharp and well defined interfaces to other materials.
Both requirements are addressed by the growth of CuMnSb thin films by molecular beam epitaxy (MBE).
This aproach additionally offers the possibility of combining CuMnSb with its ferromagnetic counterpart, the half-metallic NiMnSb, in epitaxial heterostructures.

Studying MBE grown CuMnSb can also help to obtain a better understanding of the material itself and to clarify certain discrepancies amongst material parameters found in the literature.
Since the first investigation of CuMnSb by Endo \textit{et al.} in 1968, a large range of material parameters have been collected by various research groups.
For the N\'{e}el temperature $T_{\text{N}}$, the Curie-Weiss temperature $\Theta_{\text{CW}}$ and the effective moment values ranging between \SIrange[range-phrase={{ and }}]{50}{62}{\kelvin}, \SIrange[range-phrase={{ and }}]{-250}{-120}{\kelvin} and \SIrange[range-phrase={{ and }}]{3.9}{6.3}{\mu_{B}/\unitformula}, respectively, have been reported \cite{Endo1968, Forster1968, Endo1970, Helmholdt1984, Beuf2006, Regnat2018}.
Recent calculations by M\'{a}ca \textit{et al.} suggest that the magnetic ground state of CuMnSb is determined by defects in the crystal \cite{Maca2016}.
This may explain the distribution in published values for the material parameters.

The N\'{e}el temperature of CuMnSb is accessible using standard helium cryostats.
This allows investigations of the magnetic phase transition, and comparative measurements above and below $T_{\text{N}}$ can be used to unequivocally identify which observations can be attributed to the nature of the magnetic state.

CuMnSb crystallizes in the $\text{C1}_{\text{b}}$ half-Heusler crystal structure, which can be described as four interpenetrating fcc sublattices translated by \nicefrac{1}{4} along the long diagonal of the unit cell.
One of the sublattices is unoccupied, in contrast to full-Heusler materials where all sublattices are occupied.
More details on the crystal structure and a general overview on Heusler materials can be found in \cite{Graf2011}.

In this paper we report on the epitaxial growth of CuMnSb on InAs (001) and present results of structural and magnetic characterization, as well as basic transport findings.

\section{Results and discussion}

\subsection{Epitaxial growth of CuMnSb thin films}

CuMnSb films are grown on undoped InAs (001) substrates and buffers and are subsequently capped by a \SI{5}{\nano\metre} Ru layer, by using DC sputter deposition, to protect the antiferromagnetic layer from the environment.
The epitaxial growth, as well as the deposition of the cap layer are performed in a UHV cluster system comprised of individual growth chambers connected through a UHV transfer system.

The epi-ready InAs substrate is first overgrown with an InAs buffer layer following the method given in \cite{Ye2013} using an As:In flux ratio of 15:1 and a substrate growth temperature of \SI{490}{\degreeCelsius}.
The As is supplied by a valved cracker cell, the In by a dual filament effusion cell.

The sample is then transferred to the CuMnSb growth chamber and heated up to the growth temperature of \SI{250}{\degreeCelsius}.
To stabilize material fluxes, the cell shutters are opened \SI{5}{\minute} before the growth is initiated by opening the main shutter.
Single filament effusion cells are used as sources for Mn and Sb.
Cu flux is provided by a dual filament effusion cell.
The purities of the source materials are 6N for Cu and Sb and 5N8 for Mn.

\begin{table}[b]
\caption{Optimized growth parameters resulting in the narrowest FWHM of the rocking curve and the lowest root mean square surface roughness. The beam equivalent pressures (BEP) are measured using a Bayard-Alpert type ionization gauge.}
	\centering
	\begin{ruledtabular}
	\begin{tabular}{lcc}
		Material & T (\si{\degreeCelsius}) & BEP (\si{\milli\bar}) \\
		\hline
		Cu & Tip: 1180 / Base: 996 & \num{5.5e-9} \\
		Mn & 815 & \num{8.7e-9} \\
		Sb & 424.8 & \num{4.1e-8} \\
	\end{tabular}
	\end{ruledtabular}
	\label{GROWTH_PARAMETERS}
\end{table}

\begin{figure}[tb]
	\centering
		\includegraphics{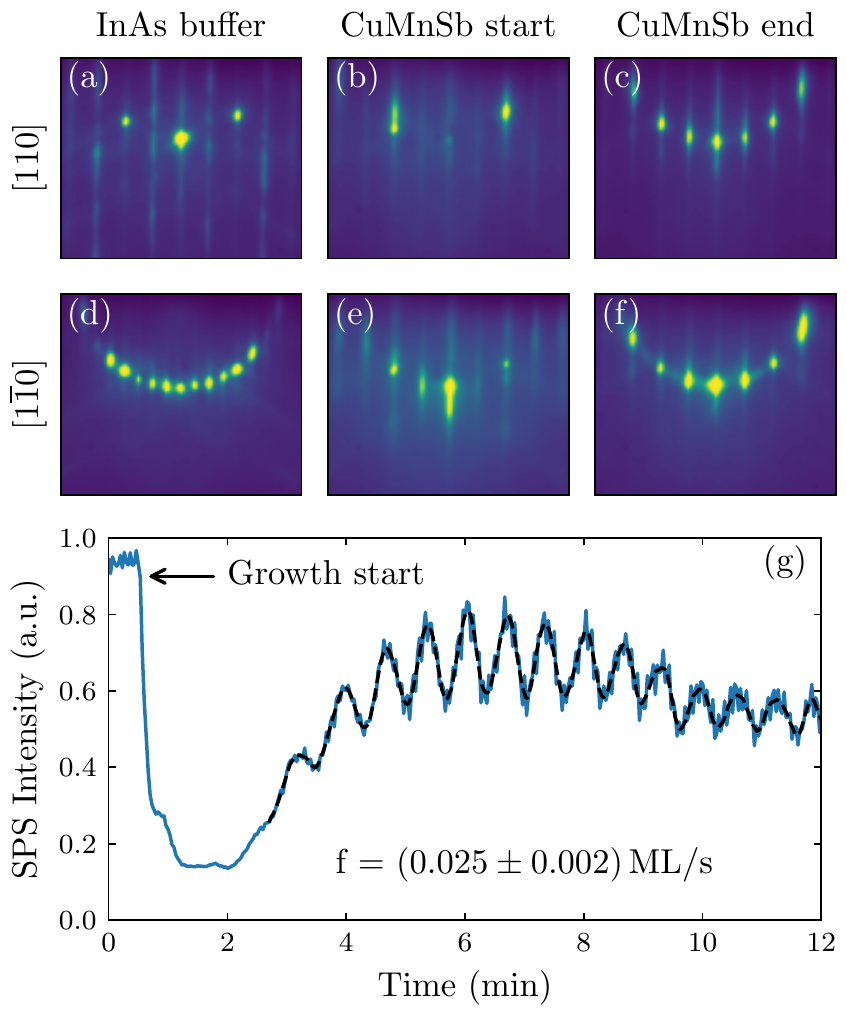}
	\caption{
		Evolution of the RHEED reconstructions during growth of CuMnSb along [110] and [1$\overline{\mbox{1}}$0] crystal directions.
		Before CuMnSb growth, an InAs $2\times4$ reconstruction is observed [(a), (d)].
		This transforms to a blurry $2\times2$ reconstruction at the growth start of CuMnSb [(b), (e)].
		With CuMnSb growth the $2\times2$ reconstruction becomes sharper and more intense [(c), (f)].
		Specular spot oscillations (g) with a period of half a unit cell confirm layer-by-layer growth.
	}
	\label{RHEED}
\end{figure}

Growth of the CuMnSb is monitored by reflection high-energy electron diffraction (RHEED).
\mbox{Fig.~\ref{RHEED}} shows the RHEED patterns at different stages of the growth.
Prior to growth start, the InAs buffer surface shows a typical $2\times4$ reconstruction, with clearly visible Kikuchi lines [Fig.~\ref{RHEED} (a), (d)] indicating a high degree of both bulk and surface ordering.
Within 2 minutes of the CuMnSb growth start (approx. 1 unit cell), the $2\times4$ reconstruction transforms to a blurry $2\times2$ pattern \mbox{[Fig.~\ref{RHEED} (b), (e)]}.
We attribute this to the formation process of the InAs/CuMnSb interface.
With increasing CuMnSb thickness, the $2\times2$ image sharpens and gains intensity [Fig.~\ref{RHEED} (c), (f)].
The points of enhanced intensity originate from a crossing of the Kikuchi lines with the elastic diffraction streaks \cite{AyahikoIchimiya2016}.
The surface reconstructions observed by RHEED do not change during typical growth times (\SI{90}{\minute} for \SI{40}{\nano\meter} thick films).
This indicates a 2D growth mode of the CuMnSb layer.

Layer-by-layer growth is confirmed by the observation of specular spot oscillations.
In Fig.~\ref{RHEED} (g) we show an example of the specular spot intensity evolution after growth start.
The drop of intensity at growth start is again attributed to the formation of the InAs/CuMnSb interface.
Combining this data with the layer thickness measured by X-ray diffraction, one oscillation period (1 monolayer) can be related to the growth of half a unit cell of CuMnSb.
The frequency of the oscillations yields a growth rate of \SI{0.025\pm0.002}{monolayer\per\second} or \SI{0.41\pm0.03}{atoms\per\nano\square\meter\second}.
This corresponds to an atomic flux of \SI{0.14\pm0.01}{atoms\per\nano\square\meter\second} for Cu and Mn. Note that re-evaporation can be neglected for a substrate temperature of \SI{250}{\degreeCelsius}.
For Sb, re-evaporation is possible and thus the atomic flux is likely slightly larger.

CuMnSb samples have been grown with a variety of fluxes and flux ratios of the elemental species.
We find that the growth of CuMnSb is as sensitive to material fluxes as was found previously for the related (ferromagnetic) material NiMnSb \cite{Gerhard2014}.
Consequently the cell temperatures must be precisely stabilized (typically to variations well below \SI{1}{\degreeCelsius}, corresponding to changes in beam equivalent pressure (BEP) of the order of \SI{1e-10}{\milli\bar}).

We find that small changes of the material fluxes impact the layer properties significantly.
For Cu and Mn, BEP variations of the order of \SI{1e-10}{\milli\bar} already lead to modifications of the magnetic properties of CuMnSb.
BEP deviations of more than \SI{2e-10}{\milli\bar} significantly reduce the crystal quality as indicated by the presence of 3D transmission spots in the RHEED pattern during growth.
In section \ref{subsec:CHARACTERIZATION} we show that for Sb, BEP variations of the order of \SI{2e-9}{\milli\bar} disturb the layer-by-layer growth and dots form on the surface of the film.

The growth parameters listed in Table \ref{GROWTH_PARAMETERS} reproducibly produce the lowest FWHM of the rocking curve measured by high resolution X-ray diffraction and root mean square roughness determined by atomic force microscopy.
Samples grown with these parameters are found to be stoichiometric from energy dispersive X-ray spectroscopy within the \SI{<3}{\percent} experimental error. All results in this paper (except for the transport measurements) are for samples grown with these parameters.

Samples dedicated for resistivity measurements are grown with an additional non-conducting ZnTe \mbox{($a=\SI{6.10}{\angstrom}$)} layer on the InAs layer to avoid parallel conductivity through the InAs buffer and substrate.
ZnTe growth is performed in the Te rich regime using a Zn:Te flux ratio of 1:2 at a substrate temperature of \SI{330}{\degreeCelsius}.

After MBE growth, all samples are capped by a \SI{5}{\nano\meter} Ru layer to protect the CuMnSb from environment.
Todo so, the samples are transferred to the sputtering chamber without breaking the UHV.
The cap layer is deposited using DC-magnetron sputtering at room temperature.
We use a \SI{5e-3}{\milli\bar} Ar pressure and a deposition rate of \SI{5}{\nano\meter\per\minute}.

\subsection{Structural characterization of CuMnSb thin films} \label{subsec:CHARACTERIZATION}

\begin{figure}[tb]
	\centering
	\includegraphics{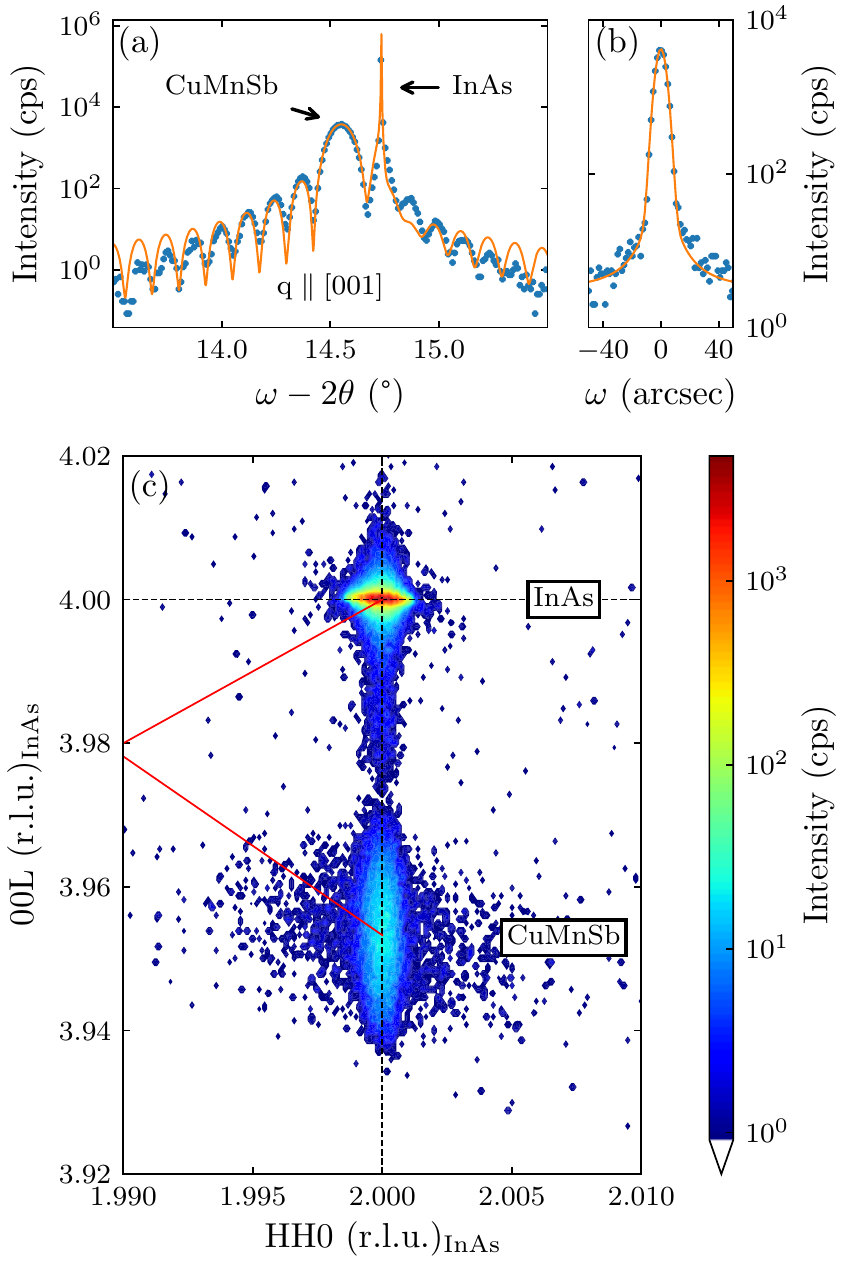}
	\caption{
		High resolution X-ray diffraction scans of the CuMnSb thin film.
		(a) $\omega - 2\theta$ diffractogram of the symmetric (002) diffraction peak (blue dots) together with a full dynamical simulation (orange line).
		(b) rocking curve of the (002) CuMnSb peak (blue dots) with the fitted Voigt profile (orange line).
		(c) Reciprocal space map of the asymmetric (224) diffraction peak.
		The relaxation triangular for the CuMnSb layer is also shown (red lines).
		The pseudomorphic CuMnSb layer shows no signs of relaxation.
	}
	\label{HRXRD}
\end{figure}

\begin{figure}[tb]
	\centering
	\includegraphics{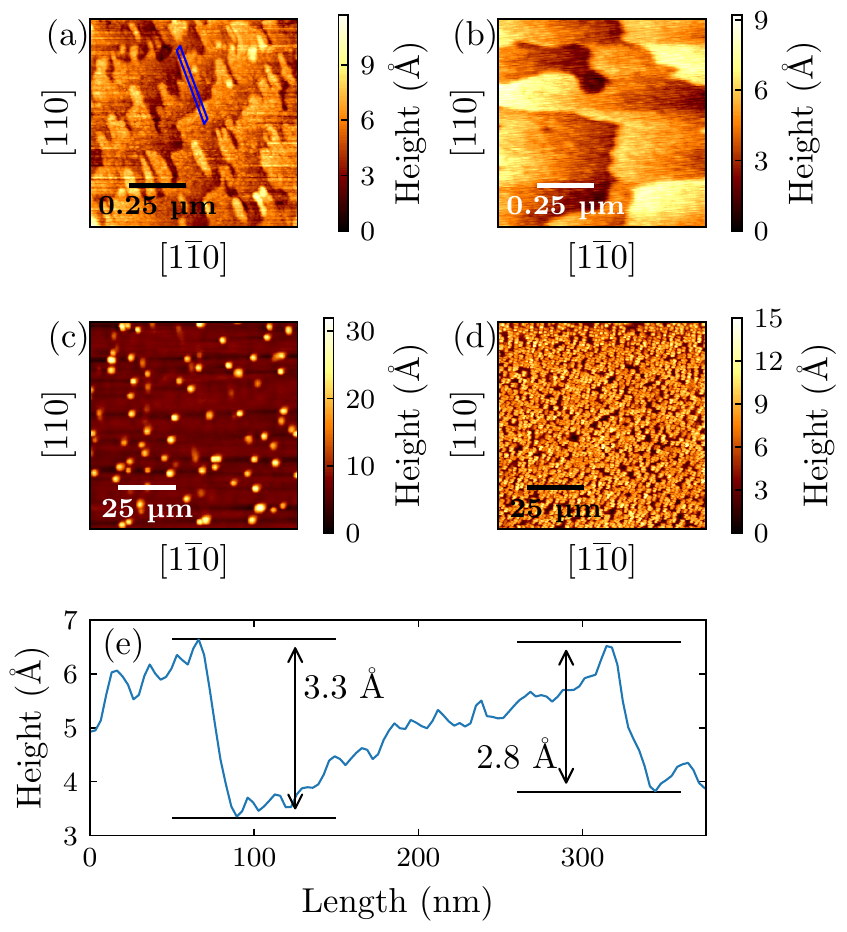}
	\caption{
		Atomic force microscopy measurements of the Ru capped CuMnSb sample (a) and an InAs buffer (b).
		The surface of Ru on CuMnSb shows atomic steps in [010] crystal direction and an root mean square roughness of \SI{0.14}{\nano\meter}.
		A decreased Sb flux of \SI{3.9e-8}{\milli\bar} leads to a formation of dots on the CuMnSb surface (c).
		Dots with a higher density are formed at the surface for an increased Sb flux of \SI{4.3e-8}{\milli\bar} (d).
		The height of the steps is consistent with half a unit cell (InAs or CuMnSb), as can be seen in the line scan (e) taken through two of the steps seen in (a).
	}
	\label{AFM}
\end{figure}

\begin{figure}[tb]
	\centering
		\includegraphics{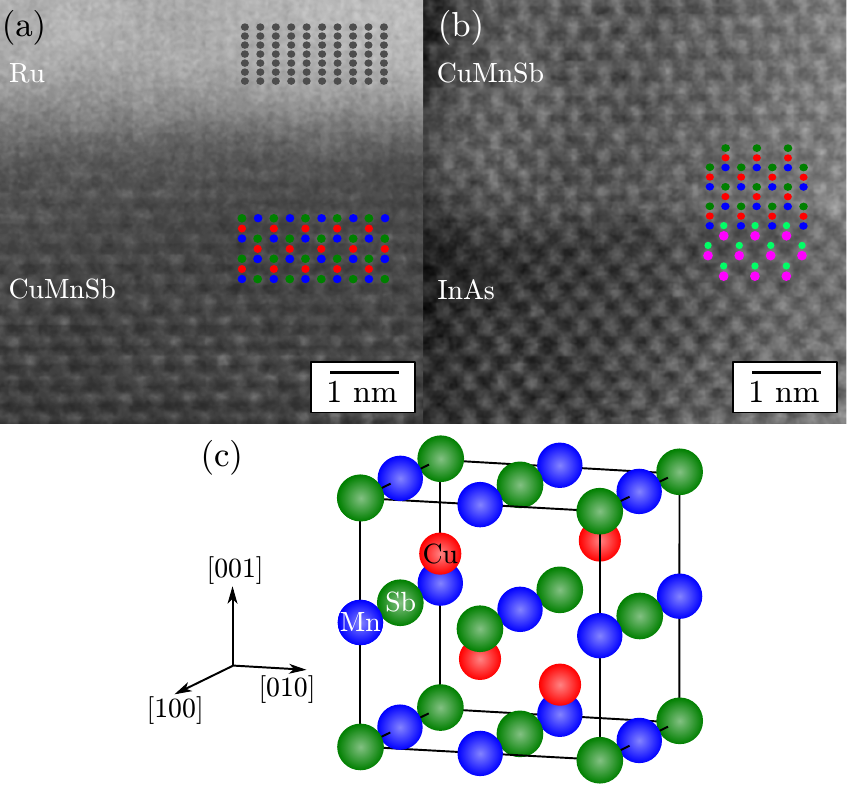}
	\caption{
		Projections of the CuMnSb and InAs crystal along the [1$\overline{\mbox{1}}$0] direction are superimposed on the STEM images [(a) and (b)].
		For this orientation, only atoms of the same species are aligned along the line of sight.
		The color assignment is as follows: Sb - dark green, Cu - red, Mn - blue, In - magenta, As - light green, Ru - gray.
    The Ru is found to grow ordered on the CuMnSb (a).
		The interface between InAs and CuMnSb (b) shows a clean transition between the two materials.
		Both layers show a high degree of order and match the overlayed crystal.
		A model of the CuMnSb unit cell is depicted in (c).
		Larger area scans can be found in the supplemental material \cite{SUP}.
	}
	\label{TEM}
\end{figure}

High resolution X-ray diffraction (HRXRD) measurements have been performed using a triple axis diffractometer.
Multiple, well-defined, thickness fringes in the $\omega - 2\theta$ diffractogram \mbox{[Fig.~\ref{HRXRD} (a)]} of the symmetric (002) peak indicate a uniform layer thickness and a high crystalline order of the \SI{40}{\nano\meter} thick CuMnSb layer.
A full dynamical simulation \cite{Kriegner2013} yields the lattice parameters.
The horizontal lattice constant is found to be \mbox{$a^{\parallel}_{\text{CuMnSb}} = a_{\text{InAs}} = \SI{6.059}{\angstrom}$}, the vertical lattice constant is found to be \mbox{$a^{\perp}_{\text{CuMnSb}} = \SI{6.136}{\angstrom}$}.
With the lattice constant of relaxed CuMnSb \mbox{$a^{relaxed}_{\text{CuMnSb}} = \SI{6.095}{\angstrom}$} \cite{Regnat2018}, this corresponds to a compressive strain of \SI{0.6}{\percent}.
Further evidence of the quality of the crystal is the \SI{7.7}{\arcsecond} FWHM of the rocking curve on the (002) peak [Fig.~\ref{HRXRD} (b)].

The pseudomorphic character of the CuMnSb layer is confirmed by a reciprocal space map of the (224) diffraction peak [Fig.~\ref{HRXRD} (c)].
The peaks corresponding to InAs and CuMnSb are aligned along the [00L] direction with no hint of broadening of the CuMnSb peak along the indicated relaxation triangle (red lines).

The surface of the CuMnSb layers are investigated by atomic force microscopy (AFM) [Fig.~\ref{AFM} (a)].
Since the Ru cap layer is deposited by sputtering it is assumed to be conformal, such that the morphology of the capped surface is a fair representation of the surface of the CuMnSb film.
The figure shows a root mean square surface roughness of \SI{0.14}{\nano\meter} over an area of \mbox{\num{1} $\times$ \SI{1}{\square\micro\meter}}.
Atomic steps with a height of half a unit cell are observed in the [010] crystal direction.
Steps can also be seen on the buffer layer [Fig.~\ref{AFM} (b)], with a lower density and no clear directionality.
The line scan in Fig.~\ref{AFM} (e) shows the height of the steps on the CuMnSb surface.
We attribute the formation of the steps to a small miscut (\SI{\pm0.1}{\degree} according to the specifications of the wafer manufacturer) of the InAs substrate.
This assignment is consistent with the spacing of the steps.

The quality of the CuMnSb film, as judged from the observed surface morphology, is significantly reduced for Sb flux deviations on the order of \SI{2e-9}{\milli\bar} BEP.
Fig.~\ref{AFM} (c) shows the surface of a CuMnSb film grown with a too low Sb flux (\SI{3.9e-8}{\milli\bar} BEP).
Randomly distributed large dots (height up to \SI{3}{\nano\meter}) with a areal density of $\sim$\SI{1}{\per\square\micro\meter} are observed.
Too high Sb flux (\SI{4.3e-8}{\milli\bar} BEP) leads to the formation of small dots with a high density of $\sim$\SI{20}{\per\square\micro\meter}.

A lamella with surface normal along the [110] crystal direction has been prepared in a focused ion beam system for imaging in STEM.
Imaging is done using a high-angle annular dark-field (HAADF) detector, and with a \SI{300}{\kilo\volt} acceleration voltage.
We find a single crystalline ordering in the epitaxial layers and a partial crystalline ordering of the Ru cap in the HAADF-STEM images [Fig.~\ref{TEM} (a) and (b)] of the Ru/CuMnSb and CuMnSb/InAs interfaces.
The mapping of the overlayed CuMnSb $\text{C1}_{\text{b}}$ crystal structure has been confirmed by a simulation of HAADF images \cite{Kirkland2010}.
The spacing of the Ru lattice planes is about \SI{2}{\angstrom}, indicating a hexagonal crystalline phase with the c-axis parallel to the interface.

The interface between InAs and CuMnSb is shown in Fig.~\ref{TEM} (b).
The arsenic atoms at the InAs surface are built into the first monolayer of CuMnSb in the position of the Sb atoms.
Larger area scans of the layerstack are shown in Fig.~S3 in the supplemental material \cite{SUP}.
No defects or stacking faults are observed.

\subsection{Magnetic measurements}

\begin{figure}[tb]
	\centering
		\includegraphics{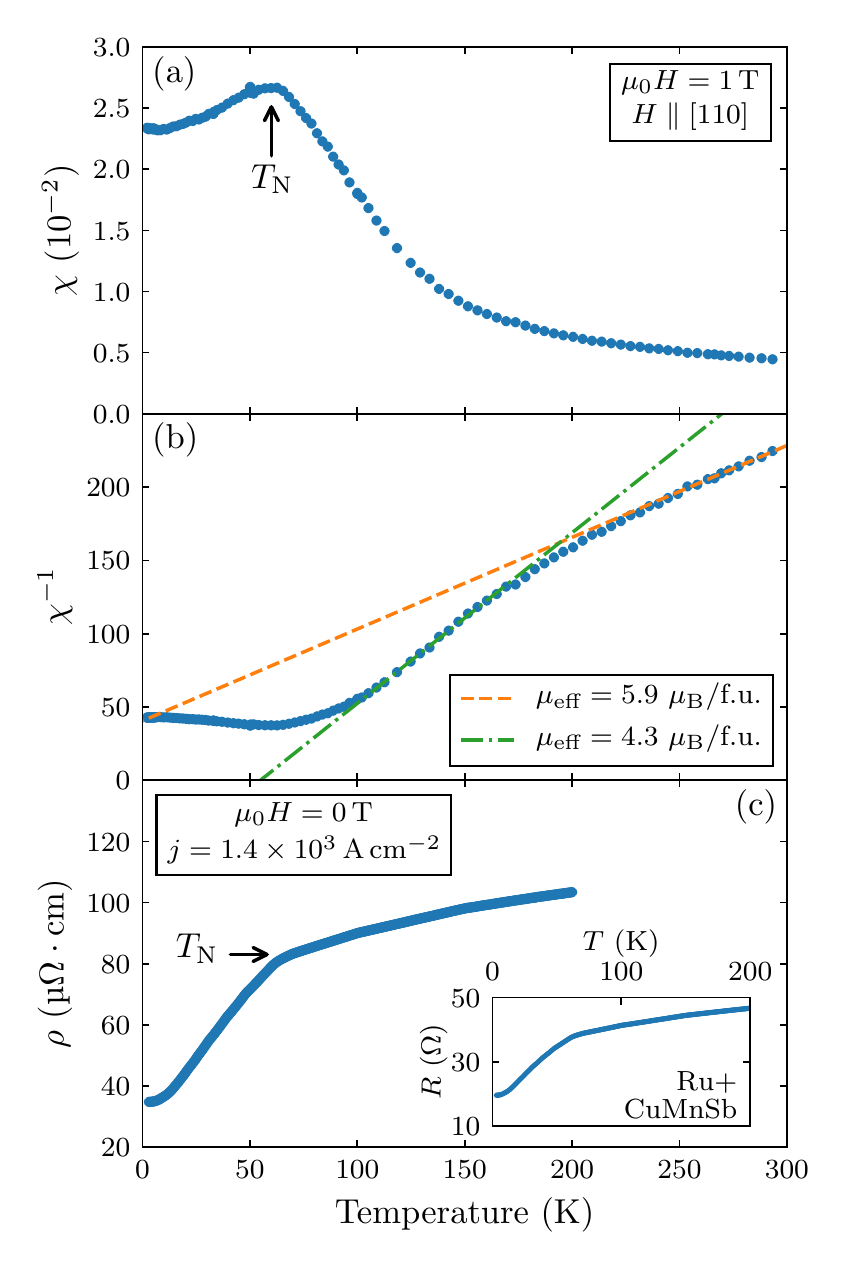}
	\caption{
		(a) Temperature dependence of the susceptibility during cooldown at \SI{1}{\tesla}.
		The cusp at \SI{62}{\kelvin} marks the transition from paramagnetic to antiferromagetic behavior ($T_{\text{N}}$).
		(b) Temperature dependent inverse susceptibility during cooldown at \SI{1}{\tesla}.
		The lines indicate the Curie-Weiss behavior in the two temperature regimes with linear behavior of the inverse susceptibility.
    (c) Temperature dependence of resistivity of the CuMnSb thin film during warmup in zero field.
    The transition from antiferromagetic to paramagnetic phase ($T_{\text{N}}$) is clearly indicated by a kink at \SI{62}{\kelvin}.
    The inset shows raw data of the sample without substraction of the comparison signal.
	}
	\label{TEMPERATURE}
\end{figure}

Magnetic characterization is performed in a commercial Superconducting Quantum Interference Device (SQUID) magnetometer using a recently designed sample holder \cite{Gas2019, Sawicki2011} to resolve and quantify the small magnetic flux of the thin antiferromagnetic CuMnSb layer by \textit{in situ} compensation of the dominant diamagnetic contribution of InAs.

The temperature dependence of the susceptibility $\chi(T)$ along the [110] crystal direction is determined in a bias field of \SI{1}{\tesla}.
Fig.~\ref{TEMPERATURE} (a) shows the $\chi(T)$ of our CuMnSb film during cooldown.
The cusp at \SI{62}{\kelvin} marks the phase transition from paramagnetism to antiferromagnetism.
The corresponding temperature is identified as the N\'{e}el temperature $T_{\text{N}}$, which is in the range of previously reported values for bulk samples (\SIrange{50}{62}{\kelvin}) \cite{Endo1968, Beuf2006, Endo1970, Helmholdt1984, Regnat2018}.

The inverse susceptibility in Fig.~\ref{TEMPERATURE} (b) shows two linear regions.
Each can be fitted to a Curie-Weiss behavior.
According to bandstructure calculations, CuMnSb carries a magnetic moment of \SI{4}{\mu_{B}/\unitformula} \cite{Jeong2005}, which results in an effective paramagnetic moment of \SI{4.9}{\mu_{B}/\unitformula}.
The fit in the temperature range above \SI{230}{\kelvin} yields an effective moment of \SI{5.9}{\mu_{\text{B}}/\unitformula} and a Curie-Weiss temperature of \SI{-65}{K}.
This is remarkably different to the Curie-Weiss temperature reported for bulk material.
A ratio of $-\Theta_{\text{CW}}/T_{\text{N}}\approx3$ is typically found in bulk samples \cite{Regnat2018}, and has been interpreted as an indication for geometric frustration \cite{Ramirez1994}.
In our layers, the ratio is close to 1.
We associate this with the epitaxial-strain induced symmetry lowering that reduces geometric frustration in the magnetic structure of the CuMnSb film.
The line fit performed in the temperature range of \SIrange{110}{170}{\kelvin} yields an effective moment of $\mu_{\text{eff}}=\SI{4.3}{\mu_{B}/\unitformula}$, which is closer to the value expected from theory \cite{Jeong2005}.
The effective moments found in both temperature regimes are within the range of reported values (\SIrange{3.9}{6.3}{\mu_{B}/\unitformula}) \cite{Beuf2006, Endo1970, Helmholdt1984, Regnat2018}.

The magnetic measurements confirm the antiferromagetic character of the CuMnSb thin films.
Extensive SQUID magnetometer studies are on-going in order to establish a more complete picture of the rich magnetic phase diagram of this material.

\subsection{Transport measurements}

Since the InAs buffer and substrate are not sufficiently insulating to allow for proper transport measurements, additional samples with a \SI{38}{\nano\metre} thick ZnTe ($a=\SI{6.10}{\angstrom}$) layer inserted between the InAs and the CuMnSb are grown.
HRXRD measurements confirm that this additional layer does not significantly impact the crystalline layer quality (see Supplemental Material \cite{SUP}).

Such a sample, with a \SI{30}{\nano\metre} thick CuMnSb layer, capped with Ru, is patterned into rectangular stripes (\num{250} $\times$ \SI{1900}{\square\micro\meter}) using standard optical lithography and physical etching.
Measurements are carried out using a 4-point geometry, DC current and a standard helium bath cryostat with a variable temperature insert.
The contacts for voltage measurement are \SI{500}{\micro\metre} apart.
A reference sample without any CuMnSb layer was also studied in order to determine the conductance of the \SI{5}{\nano\meter} Ru cap.
The resistance of the CuMnSb/Ru stack versus temperature is shown as an inset in Fig.~\ref{TEMPERATURE} (c).
This resistance is converted to a resistivity after substracting the parallel conductance from the Ru cap (see Supplemental Material \cite{SUP} for details), and shown in the main panel of Fig.~\ref{TEMPERATURE} (c).

The temperature dependent resistivity shows a clear kink at \SI{\sim62}{\kelvin}, which we associate with suppression of spin disorder scattering \cite{Mott1964} at the onset of antiferromagnetic ordering. This value for the N\'{e}el temperature is consistent with the ordering temperature found in the magnetic studies.
The residual resistivity of the sample at \SI{4}{\kelvin} of \SI{35}{\micro\ohm\cdot\centi\meter} is below the lowest values reported to date for macroscopic single crystals \cite{Regnat2018}.

\section{Conclusion}

We report on the growth of CuMnSb thin films by molecular beam epitaxy.
Crystal quality is confirmed by structural investigations and magnetic measurements find the N\'{e}el temperature around \SI{62}{\kelvin}.
Two regimes with Curie-Weiss behavior are found at temperatures above $T_{\text{N}}$.
The magnitudes of $T_{\text{N}}$, and $\mu_{\text{eff}}$ established here are in agreement with the values reported for bulk samples.
A noteworthy distinction however is the Curie-Weiss temperature which is a factor of three smaller than that of bulk samples.
The main differences to bulk samples are the crystal quality and the epitaxially induced strain.

We have also shown that lateral transport experiments can be made possible by introducing a non-conducting ZnTe layer.
This allows applications in antiferromagetic spintronics for CuMnSb and \mbox{CuMnSb/NiMnSb} heterostructures.

\begin{acknowledgments}
We thank Prof. G.~Karczewski for fruitful discussions and M.~Zipf and V.~Hock for technical assistance.
This work is funded by the Deutsche Forschungsgemeinschaft (DFG, German Research Foundation) - 397861849.
L.~Scheffler, S.~Banik, C.~Schumacher, J.~Kleinlein and L.~W.~Molenkamp acknowledge financial support by the Deutsche Forschungsgemeinschaft (DFG, German Research Foundation) in the Leibniz Program and in the projects SFB 1170 (project‐id 258499086) and SPP 1666 (project‐id 220179758), from the EU ERC Advanced Grant "4-Tops", and from the Free State of Bavaria (Elitenetzwerk Bayern IDK "Topologische Isolatoren" and the Institute for Topological Insulators).
H.~Lin and L.~W.~Molenkamp acknowledge financial support by the Deutsche Forschungsgemeinschaft (DFG, German Research Foundation) under Germany's Excellence Strategy–EXC2147 "ct.qmat" (project‐id 390858490).
\end{acknowledgments}

\end{document}


\title{Supplemental Material for Molecular beam epitaxy of the half-Heusler antiferromagnet CuMnSb}

\author{L.~Scheffler}
\affiliation{Physikalisches Institut (EP3), Universit\"at W\"urzburg, 97074 W\"urzburg, Germany}
\affiliation{Institute for Topological Insulators,  Universit\"at W\"urzburg, 97074 W\"urzburg, Germany}
\author{K.~Gas}
\affiliation{Institute of Physics, Polish Academy of Sciences, Aleja Lotnikow 32/46, 02668 Warszawa, Poland}
\author{S.~Banik}
\affiliation{Physikalisches Institut (EP3), Universit\"at W\"urzburg, 97074 W\"urzburg, Germany}
\affiliation{Institute for Topological Insulators,  Universit\"at W\"urzburg, 97074 W\"urzburg, Germany}
\author{M.~Kamp}
\affiliation{Physikalisches Institut and R\"ontgen Center for Complex Material Systems, Universit\"at W\"urzburg, 97074 W\"urzburg, Germany}
\author{J.~Knobel}
\affiliation{Physikalisches Institut (EP3), Universit\"at W\"urzburg, 97074 W\"urzburg, Germany}
\affiliation{Institute for Topological Insulators,  Universit\"at W\"urzburg, 97074 W\"urzburg, Germany}
\author{H.~Lin}
\affiliation{Physikalisches Institut (EP3), Universit\"at W\"urzburg, 97074 W\"urzburg, Germany}
\affiliation{Max-Planck-Institut f\"ur Chemische Physik fester Stoffe, 01187 Dresden, Germany}
\author{C.~Schumacher}
\affiliation{Physikalisches Institut (EP3), Universit\"at W\"urzburg, 97074 W\"urzburg, Germany}
\affiliation{Institute for Topological Insulators,  Universit\"at W\"urzburg, 97074 W\"urzburg, Germany}
\author{C.~Gould}
\affiliation{Physikalisches Institut (EP3), Universit\"at W\"urzburg, 97074 W\"urzburg, Germany}
\affiliation{Institute for Topological Insulators,  Universit\"at W\"urzburg, 97074 W\"urzburg, Germany}
\author{M.~Sawicki}
\affiliation{Institute of Physics, Polish Academy of Sciences, Aleja Lotnikow 32/46, 02668 Warszawa, Poland}
\author{J.~Kleinlein}
\email{johannes.kleinlein@physik.uni-wuerzburg.de}
\affiliation{Physikalisches Institut (EP3), Universit\"at W\"urzburg, 97074 W\"urzburg, Germany}
\affiliation{Institute for Topological Insulators,  Universit\"at W\"urzburg, 97074 W\"urzburg, Germany}
\author{L.~W.~Molenkamp}
\affiliation{Physikalisches Institut (EP3), Universit\"at W\"urzburg, 97074 W\"urzburg, Germany}
\affiliation{Institute for Topological Insulators,  Universit\"at W\"urzburg, 97074 W\"urzburg, Germany}
\affiliation{Max-Planck-Institut f\"ur Chemische Physik fester Stoffe, 01187 Dresden, Germany}

\maketitle

\begin{figure}[tb]
	\centering
		\includegraphics{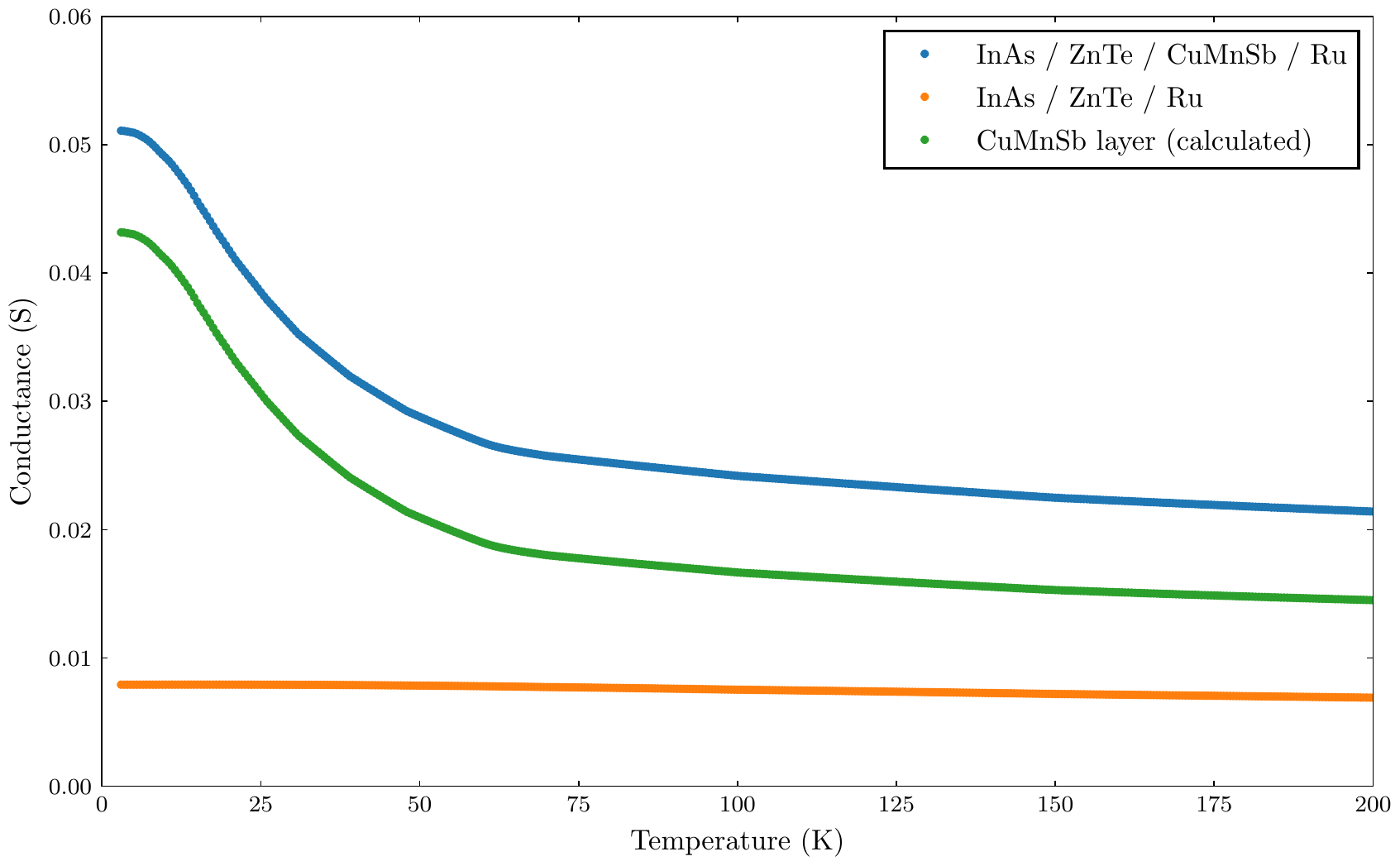}
	\caption{
		Measured conductance of the sample with the CuMnSb layer and the sample without the CuMnSb layer together with the calculate conductance of the CuMnSb layer. The conductance of the CuMnSb layer is calculated by substraction of the two measured signals. The substraction of the comparison data does not add new features to the signal.
	}
	\label{Conductance}
\end{figure}

For electrical transport measurements a \SI{30}{\nano\meter} thick CuMnSb layer is grown on a \SI{38}{\nano\metre} thick ZnTe buffer layer.
The sample is capped with a \SI{5}{\nano\metre} thick Ru layer.
To substract the effects of current shunting in this cap layer, the measurements are also performed on a comparison sample without the CuMnSb layer.
To calculate the resistivity of the CuMnSb layer, the conductance of the comparison sample is substracted from the conductance of the layerstack with CuMnSb.
This procedure is depicted in Figure \ref{Conductance}.
The substraction of the parallel conductance of the comparison sample does not add any features to the signal.

\newpage

\begin{figure}[tb]
	\centering
		\includegraphics{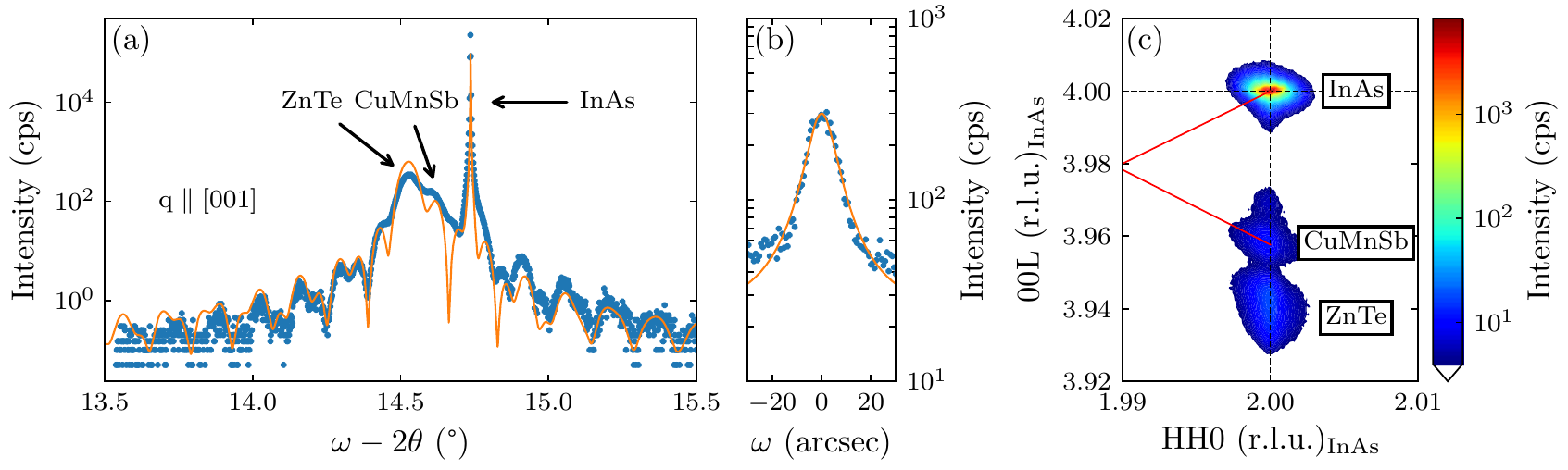}
	\caption{
	High resolution X-ray diffraction scans of a \SI{30}{\nano\meter} thick CuMnSb layer grown on a \SI{38}{\nano\metre} thick ZnTe buffer layer.
	(a) $\omega - 2\theta$ diffractogram of the symmetric (002) diffraction peak (blue dots) together with a full dynamical simulation (orange line).
	(b) rocking curve of the (002) CuMnSb peak (blue dots) with the fitted Voigt profile (orange line).
	(c) Reciprocal space map of the asymmetric (224) diffraction peak.
	The relaxation triangular for the CuMnSb layer is also shown (red lines).
	}
	\label{XRD_ZNTE}
\end{figure}

High resolution X-ray diffraction scans of a \SI{30}{\nano\meter} thick CuMnSb layer grown on \SI{38}{\nano\metre} ZnTe (Fig. \ref{XRD_ZNTE}) confirm that the additional ZnTe buffer layer does not significantly impact the CuMnSb layer quality.
The full dynamical simulation of the $\omega - 2\theta$ diffractogram [Fig. \ref{XRD_ZNTE} (a)] is used to determine the lattice parameters of the thin films.
It yields an horizontal lattice constant of \mbox{$a^{\parallel}_{\text{CuMnSb}} = a^{\parallel}_{\text{ZnTe}} = a_{\text{InAs}} = \SI{6.059}{\angstrom}$}.
The vertical lattice constant of the ZnTe buffer layer is found to be \mbox{$a^{\perp}_{\text{ZnTe}} = \SI{6.150}{\angstrom}$}, what results in a compressive strain of \SI{0.8}{\percent}.
For the CuMnSb layer a vertical lattice constant of \mbox{$a^{\perp}_{\text{CuMnSb}} = \SI{6.139}{\angstrom}$} is determined.
So the CuMnSb is compressively strained by \SI{0.7}{\percent}.
The rocking curve of the CuMnSb layer [Fig. \ref{XRD_ZNTE} (b)] shows a FWHM of \SI{13.3}{\arcsecond}.
The reciprocal space map in Fig. \ref{XRD_ZNTE} (c) confirms the full pseudomorphic character of the layerstack.

\newpage

\begin{figure}[tb]
	\centering
		\includegraphics{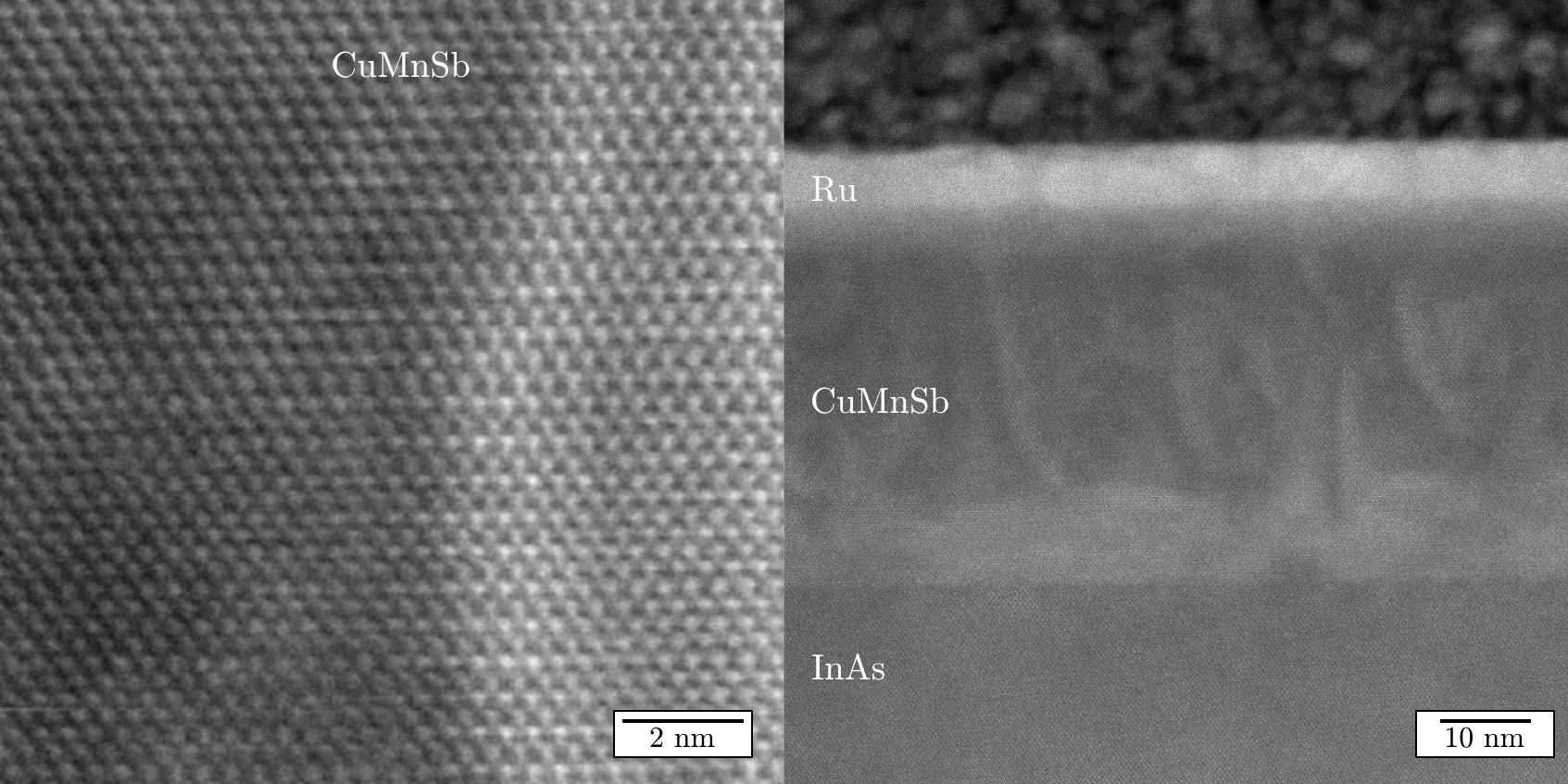}
	\caption{
	Lower magnification STEM images of the CuMnSb film grown on InAs along the [1$\overline{\mbox{1}}$0] crystal direction.
	For this orientation, only atoms of the same species are aligned along the line of sight.
	The TEM lamella is very thin, allowing the CuMnSb film to partially relax its strain by buckling.
	}
	\label{TEM_SUP}
\end{figure}